\documentclass[prd,preprint,nofootinbib]{revtex4-1} 

\usepackage{latexsym,verbatim}
\usepackage{amssymb}
\usepackage{amsfonts}
\usepackage{amsmath,color}
\usepackage{graphicx}
\usepackage{bm}
\usepackage[utf8]{inputenc}
\usepackage[T1]{fontenc}

\usepackage{hyperref}

\hypersetup{
    colorlinks=true,
    linkcolor=red,
    citecolor=green,
    filecolor=magenta,      
    urlcolor=cyan,
    pdftitle={On the low-energy limit of  stationary and axisymmetric solutions in General Relativity},
    pdfpagemode=FullScreen,}

\def\ber{\begin{eqnarray}}
\def\eer{\end{eqnarray}}
\def\beq{\begin{equation}}
\def\eeq{\end{equation}}

\def\ed{\end{document}}

\makeatletter
\let\jnl@style=\rm
\def\ref@jnl#1{{\jnl@style#1}}

\def\aj{\ref@jnl{AJ}}                   
\def\actaa{\ref@jnl{Acta Astron.}}      
\def\araa{\ref@jnl{ARA\&A}}             
\def\apj{\ref@jnl{ApJ}}                 
\def\apjl{\ref@jnl{ApJ}}                
\def\apjs{\ref@jnl{ApJS}}               
\def\ao{\ref@jnl{Appl.~Opt.}}           
\def\apss{\ref@jnl{Ap\&SS}}             
\def\aap{\ref@jnl{A\&A}}                
\def\aapr{\ref@jnl{A\&A~Rev.}}          
\def\aaps{\ref@jnl{A\&AS}}              
\def\azh{\ref@jnl{AZh}}                 
\def\baas{\ref@jnl{BAAS}}               
\def\bac{\ref@jnl{Bull. astr. Inst. Czechosl.}}
\def\caa{\ref@jnl{Chinese Astron. Astrophys.}}
\def\cjaa{\ref@jnl{Chinese J. Astron. Astrophys.}}
\def\icarus{\ref@jnl{Icarus}}           
\def\jcap{\ref@jnl{J. Cosmology Astropart. Phys.}}
\def\jrasc{\ref@jnl{JRASC}}             
\def\memras{\ref@jnl{MmRAS}}            
\def\mnras{\ref@jnl{MNRAS}}             
\def\na{\ref@jnl{New A}}                
\def\nar{\ref@jnl{New A Rev.}}          
\def\pra{\ref@jnl{Phys.~Rev.~A}}        
\def\prb{\ref@jnl{Phys.~Rev.~B}}        
\def\prc{\ref@jnl{Phys.~Rev.~C}}        
\def\prd{\ref@jnl{Phys.~Rev.~D}}        
\def\pre{\ref@jnl{Phys.~Rev.~E}}        
\def\prl{\ref@jnl{Phys.~Rev.~Lett.}}    
\def\pasa{\ref@jnl{PASA}}               
\def\pasp{\ref@jnl{PASP}}               
\def\pasj{\ref@jnl{PASJ}}               
\def\rmxaa{\ref@jnl{Rev. Mexicana Astron. Astrofis.}}%
\def\qjras{\ref@jnl{QJRAS}}             
\def\skytel{\ref@jnl{S\&T}}             
\def\solphys{\ref@jnl{Sol.~Phys.}}      
\def\sovast{\ref@jnl{Soviet~Ast.}}      
\def\ssr{\ref@jnl{Space~Sci.~Rev.}}     
\def\zap{\ref@jnl{ZAp}}                 
\def\nat{\ref@jnl{Nature}}              
\def\iaucirc{\ref@jnl{IAU~Circ.}}       
\def\aplett{\ref@jnl{Astrophys.~Lett.}} 
\def\apspr{\ref@jnl{Astrophys.~Space~Phys.~Res.}}
\def\bain{\ref@jnl{Bull.~Astron.~Inst.~Netherlands}}
\def\fcp{\ref@jnl{Fund.~Cosmic~Phys.}}  
\def\gca{\ref@jnl{Geochim.~Cosmochim.~Acta}}   
\def\grl{\ref@jnl{Geophys.~Res.~Lett.}} 
\def\jcp{\ref@jnl{J.~Chem.~Phys.}}      
\def\jgr{\ref@jnl{J.~Geophys.~Res.}}    
\def\jqsrt{\ref@jnl{J.~Quant.~Spec.~Radiat.~Transf.}}
\def\memsai{\ref@jnl{Mem.~Soc.~Astron.~Italiana}}
\def\nphysa{\ref@jnl{Nucl.~Phys.~A}}   
\def\physrep{\ref@jnl{Phys.~Rep.}}   
\def\physscr{\ref@jnl{Phys.~Scr}}   
\def\planss{\ref@jnl{Planet.~Space~Sci.}}   
\def\procspie{\ref@jnl{Proc.~SPIE}}   

\makeatother

\begin{document}

\author{Davide Astesiano}
\email{davide.astesiano@venturilab.ch}
\affiliation{Science Institute, University of Iceland,
Dunhaga 3, 107 , Reykjav\'{\i}k, Iceland}
\affiliation{Venturi Space, Route du Pâqui 1, 1720 Corminboeuf, Switzerland}

\author{Matteo Luca Ruggiero}
\email{matteoluca.ruggiero@unito.it}
\affiliation{INFN - LNL , Viale dell'Universit\`a 2, 35020 Legnaro (PD), Italy}
\affiliation{Dipartimento di Matematica ``G.Peano'', Universit\`a degli studi di Torino, Via Carlo Alberto 10, 10123 Torino, Italy}

\date{\today}

\title{On the low-energy limit of  stationary and axisymmetric solutions in General Relativity}

\begin{abstract}
We study the low-energy limit of General Relativity in the presence of stationarity and axial symmetry, coupled to dust. Specifically, we demonstrate that differences between the dynamics of General Relativity and those of Newtonian gravity persist even in the weak-field and slow-motion regime. Notably, these differences are driven by dragging terms that are not necessarily small, as is typically the case in the well-known gravitomagnetic limit. To highlight this distinction,{we use  the concept of strong gravitomagnetism that we introduced in previous works.} We provide a pedagogical discussion of how these discrepancies arise and outline a systematic procedure to solve the equations of motion for such systems. Furthermore, we present analytical results for specific cases and also give the general solution for the vacuum case. A particularly notable result is our demonstration of how General Relativity can naturally account for a Tully-Fisher-like relation.
\end{abstract}

\maketitle

\section{Introduction }\label{sec:intro}

The quest for experimental tests of General Relativity (GR) in  terrestrial or solar system environment has  always been characterized by the need for highly accurate measurements to detect very small effects. In fact, with the remarkable exception of the advance of Mercury perihelion, whose order of magnitude is not different from other Newtonian corrections \cite{mercury}, classical tests of GR are usually difficult to be detected when the gravitational field is week and the speeds of the sources are small if compared to the speed of light. Nonetheless, Einstein's theory is the best description we have of gravitational interactions in our Universe: from the neighborhood of the Earth to the far reaches of space, where spacetime is shaped by the presence of very compact objects that are fast-moving or rotating and produce intense gravitational fields, GR has passed all tests with flying colors, greatly enhancing our understanding of fundamental physics \cite{willrev}. Indeed, the existence of neutron stars, black holes and  the emission of gravitational waves prove that GR gives rise to gravitational phenomena which do not have a Newtonian analogue. Additionally, also in conditions of weak gravitational field  there are GR effects with no Newtonian counterpart: this is the case of the so-called gravitomagnetic effects \cite{Ruggiero:2023ker} which are determined by mass currents: these effects give rise to non-diagonal terms in the metric tensor and are responsible for the dragging of inertial frames \cite{ciufolini1995gravitation}. Gravitomagnetic effects mark a sharp difference with Newtonian physics as a simple example may suggest: the gravitational field inside a rotating shell is not null in GR \cite{ciufoliniricci} but it is clearly zero in Newtonian physics.

The purpose of this paper is to focus on another situation where GR effects not only lack Newtonian analogue but, in addition, they are of the same order of Newtonian ones: we are talking about  self-gravitating systems, made of dust, in stationary and axially symmetric rotation.  These systems have been investigated in earlier studies \cite{Astesiano:2021ren,Astesiano:2022gph,Astesiano:2022ozl,Astesiano:2022ghr,Ruggiero:2023geh} using a top-down approach that starts from the exact solution of Einstein’s equations and leads to suitable low-energy limits. In contrast, here we adopt a bottom-up strategy that starts from the low-energy limit of the metric with the appropriate symmetries and derives the corresponding solution of Einstein's equations for different low-energy limits.    

The paper is organized as follows: after introducing in Section \ref{sec:hyp} the model of the system, in Section \ref{sec:lel} we discuss its low-energy limits and outline the equations of motion which describe the dynamics of the sources and their gravitational field. In Section \ref{sec:sol} we discuss a solution strategy of the model equations and, subsequently, in Section \ref{sec:komar} we give a physical interpretation of the system in terms of the Komar integrals corresponding to its mass and angular momentum. Example of solutions are discussed in Section \ref{sec:exsolutions}. The most general vacuum solution under these assumptions is given in Section \ref{vacuum}, while conclusions are in Section \ref{sec:conc}.

\section{The Model} \label{sec:hyp}

The system we are considering is constituted by a stationary and axially symmetric rotating dust fluid; accordingly, its energy-momentum tensor can be written as
\begin{align}
    T= \rho \, u \otimes u, \label{Dust}
\end{align}
for a fluid without pressure with four velocity $u$ and density $\rho$. The stationarity and axial symmetry imply the existence of one timelike Killing vector and one spacelike Killing vector: accordingly, Einstein's equations can be integrated up to quadratures, using techniques that can be traced to the work by \citet{Geroch:1970nt},\cite{Geroch:1972yt}, and subsequently by Hansen and Winicour  \cite{HansenWinicour1}, \cite{HansenWinicour2}, \cite{Winicour}. For an in-depth discussion, we refer to the classical book by \citet{stephani_kramer_maccallum_hoenselaers_herlt_2003}.

Given these symmetries, the physical features of a  solution are summarized by the velocity and density profiles of the sources. In addition, we remark that for this class of solutions, a very important feature is that given the velocity profile and the density profile, the solution is non-unique, but there are still some degrees of freedom, as emphasised in the paper by  \citet{Astesiano:2021ren}, where the differences between the general GR solution and the corresponding Newtonian one are discussed. Here, we are concerned with the low-energy limits of these solutions: this amounts to considering them in the weak-field regime for the gravitational field. In particular, this work is an ideal prosecution of what we did in a previous publication \cite{Astesiano:2022ozl}, where we considered a specific class of solution.

That said, we consider a region $\mathcal{R}$ of the spacetime $\mathcal{V}_4$ where  a self-gravitating system made of dust (described by Eq. (\ref{Dust})) is present. As we said, we assume that dust is rotating with axial symmetry and that out of $\mathcal{R}$ the gravitational field becomes negligibly small, so that   the solution asymptotes flat Minkowski spacetime $\mathcal{M}_{4}$. We can introduce adapted cylindrical coordinates $(x^0=ct,r,z,\phi)$, such that 
\begin{itemize}
    \item $\partial_0$ and $\partial_\phi$ are the Killing vectors mentioned above, 
    \item the matter flows along the Killing vectors $\partial_0$ and $\partial_\phi$
    \item all functions and tensors depend only on the coordinates which are not associated to Killing vectors, i.e. $(r,z)$.
\end{itemize}
The time coordinate $x^0=ct$ labels the spacelike hypersurfaces which are invariant under time translations and $\phi$ is the axial-angle coordinate around the axis of symmetry. Furthermore, the energy-momentum tensor is invariant under a simultaneous change of sign of $ct$ and $\phi$.  Accordingly, the metric-tensor is in the form (see e.g. \citet{1970ApJ...162...71B}): 
\begin{align}
    ds^2= \frac{g_{tt}}{c^2} \left(cdt\right)^{2}+ 2 \frac{g_{t\phi}}{c} \left(cdt\right) d\phi+ g_{\phi\phi} d\phi^2+g_{rr} dr^2+g_{zz} dz^2, \label{metric}
\end{align}
In the above expression $g_{\mu\nu}=g_{\mu\nu}(r,z)$; far from the region $\mathcal{R}$ where the system is located, the metric (\ref{metric}) becomes $\eta_{\mu\nu}$, which is the Minkowski metric written using cylindrical coordinates 
\beq
ds^2=- \left(cdt\right)^{2}+dr^{2}+r^{2}d\phi^{2}+dz^{2}. \label{eq:metricflat}
\eeq
In the spacetime modeled by (\ref{metric}) we can introduce different observers \cite{Bini:2008uyx}. In particular, we are interested in the static observers and in the zero angular momentum observers ``ZAMO'' (see e.g. \citet{1972ApJ}).  Static observers are, by definition, at rest with respect to the adapted coordinates; however, for this kind of systems they are not a good choice, because these observers are not defined by local properties of spacetime and, in addition, they cannot exist in some regions  (see e.g. \citet{1970ApJ...162...71B}). Accordingly, the so-called ZAMO or non-rotating observers are introduced: by definition, their angular momentum vanishes.

The static observers are defined by the four-velocity S
\begin{align}
    S= \frac{1}{\sqrt{-g_{tt}}} \partial_t.
\end{align}
Due to the assumptions on the asymptotic flatness of the metric, the reference frame defined by the  congruence of worldlines $\partial_t$ corresponds to a rigid frame anchored to the asymptotic inertial frame at infinity. It is at rest compared to the distant stars, but it has non-zero angular momentum  in the region $\mathcal{R}$  \cite{universe7100388}.

On the other hand, the ZAMO observers are defined by the request that these observers are non-rotating, in the sense that they are orthogonal to the constant time spacelike hypersurfaces $\Sigma_t$ and define a field of one-forms
\begin{gather}
    Z= \frac{r}{\sqrt{g_{\phi\phi}}} dt. \label{E0ZAMO}
\end{gather}
Despite having zero angular momentum they have non-zero angular velocity $\chi$ as measured in a reference frame fixed relative to the distant stars:
\begin{align}
    \chi = - \frac{g_{t\phi}}{g_{\phi\phi}}.
\end{align}

After describing  the relevant class of observers,  we consider the four-velocity $u$ of a fluid element, which must be a linear combination of the Killing
vectors and can be written as
\begin{align}
  u^{\mu}(r,z)= u^t(r,z) \left(1,0,0,\Omega(r,z) \right). \quad \label{eq:fluidvel}
\end{align}
The function $\Omega(r,z)= \frac{d\phi}{dt}= \frac{u^{\phi}}{u^{t}}$ is the angular velocity  seen by the static observers and, therefore, it corresponds to 
the velocity measured by the asymptotic inertial observers. The coordinate expression of the energy momentum tensor (\ref{Dust}) is given by
\begin{align}
    T^{\mu\nu} (r,z)= \rho(r,z) u^{\mu}(r,z) u^{\nu}(r,z).
\end{align}    

\section{The low-energy limit} \label{sec:lel}

After introducing the key features of the model, here we are going to carefully study its low-energy limit. As customary in this case, we assume the existence of global coordinates $(t,x^i)$ in which the metric reads
\begin{gather}   g_{\mu\nu}=\eta_{\mu\nu}+h_{\mu\nu}, \qquad \left|\frac{h_{\mu\nu}}{\eta_{\mu\nu}} \right| \ll 1. \label{expmetric}
\end{gather}

As we said above, these coordinates naturally select a preferred reference frame $\mathcal{I}$, defined by the family of curves $t=$variable, which also represents the asymptotic inertial reference frame. We assume that in this reference frame the motion of the matter content is slow if compared to the speed of light. To develop the low-energy limit, we usually perform an expansion in powers of $1/c$ and it is generally assumed that there are no other parameters in the expansion which are comparable with $c$. For example, in the Kerr spacetime the angular momentum of the source defines a parameter $a$ such that
\begin{align}
    a=\frac{J}{Mc},
\end{align}
where $M$ is the mass and $J$ the angular momentum. In the low-energy limit this parameter it is assumed to be $a\sim O(c^{-1})$. 

The expansion of the metric (\ref{metric}) is guided by the expansion in powers of $1/c$ of the matter terms through Einstein's equations
\begin{align}
    R_{\mu\nu}-\frac{1}{2}g_{\mu\nu}R= \frac{8\pi G}{c^4} \rho u_{\mu} u_{\nu},
\end{align}
with the normalisation condition
\begin{align}
    u^{\mu}u_{\mu}=-c^2.
    \label{realcond}
\end{align}
The trace reversed Einstein's equations turn out to be
\begin{align}
    R_{\mu\nu}= \frac{1}{2}g_{\mu\nu} \left(\frac{8\pi G}{c^2} \rho\right)+ \frac{8\pi G}{c^4} \left(\rho u_{\mu} u_{\nu}\right),
\end{align}
and it is useful to write them in the following form
\begin{align}
    R_{\mu\nu}- \tilde{R}_{\mu\nu}&=0, \label{EinsEq}\\
    \tilde{R}_{\mu\nu}&:= \frac{8\pi G}{c^4}\rho \left(u_{\mu} u_{\nu}+c^{2}\frac{g_{\mu\nu}}{2}\right).
\end{align}

The expansion can be written\footnote{The notation used $f=1+O\left(c^{-n}\right)$ means that in the low-energy limit considered, the given function $f$ differs from $1$ by a term of the order of $ \frac{1}{c^{n}}$.  } as:
\begin{align}
     ds^2&= -c^2\left(1+O(c^{-2})\right)dt^2+2 O(c^{-1}) c dt d\phi +r^2\left(1+O(c^{-2})\right)d\phi^2+\left(1+O(c^{-2})\right) \left(dr^2+dz^2\right), \\
     u^t&=1+O(c^{-2}), \quad u^\phi=O(c^{-0}), \quad
        u_t=-c^2+O(c^{-0}) , \quad u_\phi=O(c^{-0}).
\end{align}

The expansion of the metric is motivated by the condition expressed by Eq. (\ref{expmetric}) and by symmetry arguments, while the expansion of the dust four velocity $u$ is motivated by Eq. (\ref{realcond}).

Accordingly, the leading orders of the explicit components of $\tilde{R}_{\mu\nu}$ along the Killing vectors are
\begin{align}
    \tilde{R}_{tt}&=  \frac{8\pi G\rho}{2}+O(c^{-2}), \label{tt}\\
    \tilde{R}_{t\phi}&=0+O(c^{-2}), \label{tphi}\\
    \tilde{R}_{\phi\phi}&= \frac{8\pi G}{c^2} \frac{\rho}{2} r^2+ O(c^{-4})= \frac{1}{c^2} r^2 \tilde{R}_{tt}+O(c^{-4}) \label{phiphi}.
\end{align}
We notice that $\tilde{R}_{tt}$ is evaluated at order $O(c^{-0})$ while the term $\tilde{R}_{\phi\phi}$ at the order $O(c^{-2})$: this is because they have different dimensions. The same goes for $\tilde{R}_{t\phi}$. In fact, if we introduce the coordinate $x^0=ct$, then 
\begin{align}
    \tilde{R}_{00}&= \frac{8\pi G\rho}{2 c^2}+O(c^{-4}), \\
    \tilde{R}_{0\phi}&=0+O(c^{-3}), \label{AngEq}\\
    [\tilde{R}_{00}]&=[\frac{\tilde{R}_{0\phi}}{r}]= [\frac{\tilde{R}_{\phi\phi}}{r^2}]= \text{length}^{-2}.
\end{align}
The above results clearly show that the leading term of the Ricci curvature is of order $c^{-2}$.

Let us focus on the non-diagonal term $\tilde{R}_{t\phi}$. Eq. (\ref{AngEq}) is not forcing necessarily 
the non-diagonal metric terms  to be of higher order in $c^{-1}$ respect to the diagonal term. It is simply stating that the Ricci curvature must satisfy
\begin{align}
    \tilde R_{0\phi}=0+O(c^{-3}),
\label{DraggEq}
\end{align}
whence we get  a differential equation for $g_{t\phi}$. As discussed in the introduction, these off-diagonal terms are responsible for the dragging of inertial frames: accordingly, we will refer to them as   \textit{dragging effects} or \textit{dragging terms.}

At higher order in the expansion we get a differential equation between $g_{t\phi}$, $\rho$ and $\Omega$: 
\begin{align}
   \tilde{R}_{0\phi}=\frac{1}{c}\tilde{R}_{t\phi}= 0+ \frac{4 \pi  G \rho \left(g_{t\phi}-2 r^2 \Omega \right)}{c^3}+O(c^{-5}). \label{Highordtphi}
\end{align}
In particular, the Newtonian limit \textit{where no dragging effects are present}  is obtained assuming that the solution to $\displaystyle R_{0\phi}=0+O(c^{-3})$  at the leading order is just $\displaystyle     g_{t\phi}= 0+ O(c^{-a})$, where $a$ is positive. However,  this is not the  general case.\\

That said, the metric which describes the spacetime of the system is solution of Eq. (\ref{EinsEq}). Previously (see e.g. \citet{Astesiano:2022ozl}) this metric was obtained  from the exact  solution of Einstein's equation and subsequently performing the low-energy limit. Here, we  consider the opposite approach, starting from the low-energy limit before solving Einstein's equations. The two procedures coincide and give the following solution at the leading order
\begin{align}
     ds^2&= -c^2\left(1-\frac{2\Phi}{c^2}-\frac{\psi^2}{r^2c^2}\right)dt^2-2 \frac{\psi}{c} c dt d\phi +r^2\left(1+ \frac{2\Phi}{c^2}\right)d\phi^2+ e^{\Psi} \left(dr^2+dz^2\right), \label{MFD} \\
     u^t&=1+ \frac{\Phi}{c^2}+ \frac{v^2}{2c^2},\quad u^\phi=\Omega=\frac{\psi}{r^2}+ \frac{v}{r}, \label{eq:defv}
\end{align}

In the above Eq. (\ref{eq:defv}) $v$ is defined as $\displaystyle v= V-\frac{\psi}{r}$, where $V:= r \Omega$: in particular $v$ is the dust velocity as measured by the ZAMO (see e.g. \citet{Astesiano:2022ozl}), while $V$ is the dust velocity measured from asymptotic inertial observers. In addition, we can fix the function $e^{\Psi}$ by means of a line integral, since
\begin{align}
    \Psi_{,r}=\frac{1}{2r} \left[2r \partial_r\left(\frac{2\Phi}{c^2}+ \frac{\psi^2}{c^2r^2}\right)+ \frac{\psi^2_{,z}-\psi^2_{,r}}{c^2} \right]+ O(c^{-4}), \label{eq:defPSIr}
\end{align}
or, equivalently,
\begin{align}
 \Psi_{,z}=\frac{1}{2r} \left[2r \partial_z\left(\frac{2\Phi}{c^2}+ \frac{\psi^2}{c^2r^2}\right)-\frac{2}{c^2} \psi_{,r} \psi_{,z}\right]+ O(c^{-4}). \label{eq:defPSIz}
\end{align}
However, these functions are not needed in the following discussion. \\

To discuss the physical meaning of the solution (\ref{MFD}) and its peculiar non-Newtonian features, we focus on the motion of the dust particles that are sources of the gravitational field. To this end, from the metric (\ref{MFD}) we may write the Lagrangian 
\beq
L= \frac 1 2 g_{\mu\nu} \dot x^{\mu}\dot x^{\nu}, \label{eq:lagr1}
\eeq
where dot means derivation with respect to the coordinate time $t$, which we use to parameterise the test particle world-lines. Due to the symmetries of the system, we have the conserved quantity
\beq
\frac{\partial  L}{\partial \dot \phi}=p_{\phi} \label{eq:pphi}
\eeq 
that, up to the required order, can be written as
\beq
p_{\phi}=r\left(V-\frac{\psi}{r}\right)=rv \label{eq:pphi1}
\eeq
and can be interpreted as the angular momentum for unit mass. In addition, $L$ is a conserved quantity  along the geodesics of the system; in particular, due to the symmetry of the system, we are interested in the circular geodesics ($r=const$) in planes parallel to the equatorial plane ($z=const$), where $ L$ can be written as
\beq
\Phi+\frac{1}{2}\frac{\psi^{2}}{r^{2}}-\frac{\psi V}{r}+\frac{V^{2}}{2}=\mathrm{const}. \label{eq:Lconst}
\eeq
To obtain the geodesics, we may write the Euler-Lagrange equations
\begin{eqnarray}
\frac{d}{dt}\frac{\partial  L}{\partial \dot z} &=& \frac{\partial  L}{\partial z}, \label{eq:ELz} \\
\frac{d}{dt}\frac{\partial  L}{\partial \dot r} &=& \frac{\partial  L}{\partial r},  \label{eq:ELr}
\end{eqnarray}
and set, respectively, $\dot z=0$, $\dot r=0$. The geodesic equations together with  Einstein's equations at the leading order read:
\begin{align}
     \frac{V}{r} \partial_z \psi&=\partial_z \Phi+ \frac{\psi}{r^2} \partial_z \psi, \label{SGMzz}\\
     \frac{V}{r} \partial_r \psi&=\partial_r \Phi+  \frac{V^2}{r}+ \frac{\psi}{r} \partial_r \frac{\psi}{r},\label{SGMrr}\\
   \partial_{rr} \psi+ \partial_{zz} \psi- \frac{\partial_r \psi}{r}&=0, \label{Sgmsource1}\\
   \rho&= -\frac{1}{4\pi G} \left[\nabla^2 \Phi+ \frac{(\partial_{z}\psi)^2+(\partial_r\psi-2 \frac{\psi}{r})^2}{2 r^2}\right]\label{Sgmsource2}.
\end{align}
The generalization to the case of a perfect fluid where the pressure is not negligible is given in Appendix (\ref{Addition of pressure}). {A slightly different approach to the low-energy limit for stationary axisymmetric rotating solutions, with internal pressure, was considered by \citet{Galoppo:2024ttc}.}

We point out that Eq.  (\ref{Sgmsource1}) is simply Eq. (\ref{DraggEq}) for the angular part of the Ricci tensor and it is in the form of a homogenous Grad-Shafranov equation \cite{grad1958hydromagnetic,shafranov1958magnetohydrodynamical}.

We remark that no $1/c$ terms are present in these equations, which shows that the leading order is intrinsically \textit{non-Newtonian.}  
 As we already discussed, despite $\psi=0+O(c^{-1})$ is a solution, it is not the general solution, and there are solutions for this dragging term that are of the same order as the gravitational potential $\Phi$.  This is a very important point, which shows that \textit{in the low-energy limit the discrepancy between GR and Newtonian dynamics does not vanish.}  We point out that these solutions are needed to guarantee the existence of the system: in fact, Eq. (\ref{SGMzz}) suggests that to have an equilibrium along the symmetry axis, the rotation effects determined by  $\psi$  and deriving from the off-diagonals terms in the spacetime metric cannot be negligible with respect to the Newtonian ones, represented by $\Phi$.

In addition, we see that the Newtonian limit is restored when $\psi=0+O(c^{-2})$. In fact, Eq. (\ref{SGMrr}) and Eq. (\ref{Sgmsource2})  at the leading order become the usual condition for circular orbits in Newtonian dynamics and the Poisson equation, respectively.\\

The equation for the density (\ref{Sgmsource2}) is very interesting. We see that for fixed $V$, the matter density of the Newtonian case (when $\psi\rightarrow 0 $) is always greater compared to the case where $\psi$ is significant. Therefore, the presence of $\psi$ is reducing the density required to sustain the motion compared to Newtonian dynamics. This information is in the term
\begin{align}
    \rho_{(\psi^2)}:= \frac{1}{4\pi G} \frac{(\partial_{z}\psi)^2+(\partial_r\psi-2 \frac{\psi}{r})^2}{2 r^2} \label{eq:rhopsi}
\end{align}
which can be thought of as an effective density determined by the dragging term $\psi$. The presence of this additional density modifies the interplay between the sources of the gravitational field and the Newtonian potential $\Phi$:  the key point is that these additional sources (\ref{eq:rhopsi}) are not necessary small, as we have seen.{This coupling between the dragging term $\psi$ and the Newtonian potential $\Phi$ can be also testified by a scalar-vector-tensor analysis of the metric (\ref{MFD}) as we show in Appendix \ref{sec:scalar}: in fact, the scalar degree of freedom contains both $\Phi$ and $\psi$, which is not the case in the standard gravitomagnetic limit. We suggest that this arises as a consequence of the symmetries of the system, which, as we discussed above, is ultimately substained by its rotation.}
These ``new solutions'' for the $g_{t\phi}$ term are not directly related to mass currents and are responsible for what we called \textit{strong gravitomagnentism} in the previous paper \cite{Astesiano:2022ozl}.  {We note that, starting from the same exact solution considered in \cite{Astesiano:2022ozl}, \citet{Re:2024qco} showed that a non-linear gravitoelectromagnetic formalism can be obtained.}

Another interesting limit that we can reproduce is the standard gravitomagnetic limit (see e.g. \citet{Ruggiero:2002hz,Mashhoon:2003ax,Ruggiero:2023ker}). To obtain it, we must assume
\begin{align}
   \psi=0+O(c^{-1}),
\end{align}
When we neglect{non linear terms in $\psi$ and its derivatives,} we obtain the gravitomagnetic limit,  which exactly means that we are rescaling $\psi \rightarrow \psi/c$. The equations corresponding to (\ref{SGMzz})-(\ref{Sgmsource2}) are 
\begin{align}
\frac{V}{r}\partial_z \psi&= \partial_z\Phi \label{0emz}, \\ 
    V^2&=r(-\partial_r\Phi)+ V \partial_r\psi, \label{0emr} \\
    \partial_{zz} \psi+\partial_{rr} \psi- \frac{\partial_r \psi}{r}& \label{0GMA}=\frac{4\pi G}{c} \rho V, \\
      \partial_{zz} \Phi+\partial_{rr} \Phi+ \frac{\partial_r \Phi}{r}&= -4\pi G \rho \label{0PoissonGM}.
\end{align}
(see also \citet{Astesiano:2022ozl,Astesiano:2022ghr}).  

A confrontation between the set of equations (\ref{SGMzz})-(\ref{Sgmsource2}) defining  strong gravitomagnetism and the above ones, defining standard gravitomagnetism, suggests a key difference which stems from the equation for the dragging term $\psi$. In fact, while in the first case $\psi$ is the solution of the homogenous equation (\ref{Sgmsource1}), in the second case it is related to the sources through the non-homogenous equation (\ref{0GMA}): as a consequence, in standard gravitomagnetism the dragging terms are usually dumped by a $V/c$ term.  

In what follows we investigate the peculiarities of the strong gravitomagnetic limit.

\section{Integration of the strong gravitomagnetic equations} \label{sec:sol}

In this Section we are concerned with the solution of the equations of motion of the low-energy limit (\ref{SGMzz})-(\ref{Sgmsource2}) which define what we called strong gravitomagnetic limit.  To begin with, we see that the geodesic equations (\ref{SGMzz}) and (\ref{SGMrr}) can be seen just as  condition for the existence of a function $\Phi$. Using $v= V-\frac{\psi}{r}$,  Eqs. (\ref{SGMzz}) and (\ref{SGMrr}) can be rewritten as
\begin{align}
    d\Phi= -\frac{v}{r^2} \left(2\psi+r v\right) dr+ \frac{v}{r} d\psi.
\end{align}
We can rewrite this equation as
\begin{align}
   d\Phi= -\frac{p_{\phi}^2}{r^3} dr+ p_{\phi} d\left(\frac{\psi}{r^2}\right), \label{eq:dPhi1}
\end{align}
where $p_{\phi}$ is the angular momentum per unit mass $p_\phi= rv$, given in Eq. (\ref{eq:pphi1}). Notice that equation (\ref{eq:dPhi1}) is equivalent to the constraint (\ref{eq:Lconst}).  The condition for the existence of the function $\Phi$ is therefore $\displaystyle d (d\Phi)=0$. Using the relation
\begin{align}
    \psi= rV-r v= \Omega r^2-p_\phi,
\end{align}
we end up with
\begin{align}
    d \left[ -\frac{p_{\phi}^2}{r^3} dr+ p_{\phi} d\left(\Omega-\frac{p_\phi}{r^2}\right) \right]=0.
\end{align}
We deduce that $p_\phi$ is a function of $\Omega$ only. In fact, after imposing $p_\phi=p_\phi(\Omega)${and using a prime ($'$) to indicate derivative with respect to $\Omega$,} we get
\begin{align}
    d \left[ -\frac{p_{\phi}^2}{r^3} dr+ p_{\phi} d\left(\Omega-\frac{p_\phi}{r^2}\right) \right]=& d \left[\frac{p_{\phi}^2}{r^3} dr+ \left(p_{\phi} -\frac{ p_\phi p'_\phi}{r^2} \right) d\Omega \right]= \nonumber\\
    =& d\left[d\left(- \frac{1}{2}\frac{p_\phi^2}{r^2}+\int p_\phi d\Omega\right)\right]=0
    .
\end{align}
This also tells us that the Newtonian potential fulfills the following equation
\begin{align}
    \Phi= - \frac{1}{2}\frac{p_\phi^2}{r^2}+\int p_\phi d\Omega+\text{constants}. \label{Newtpot}
\end{align}
 Therefore, Eq. (\ref{Newtpot}) and
\begin{align}
    \psi= r^2 \Omega -p_\phi(\Omega), \label{SolV}
\end{align}
represent the solution of the geodesic equations, where $r v= p_\phi$ is a generic function of $\Omega$.

Then, Eqs. (\ref{SGMrr}) and (\ref{SGMzz}) are integrated and we end up with two equations and two unknowns, which are $\Omega$ and $\rho$. 
Inserting the above relations in Eqs. (\ref{Sgmsource1}) and (\ref{Sgmsource2}) we get
\begin{align}
    \rho=&\frac{ (\partial_r\Omega) ^2+(\partial_z\Omega) ^2}{8 \pi  G r^2}
   \left(p_\phi'(\Omega)^2-r^4\right), \label{densityred} \\
    0=&\left[(\partial_r\Omega)^2+(\partial_z\Omega)^2\right] p_\phi''(\Omega)+ \left(p_\phi'(\Omega)+r^2\right) \left(\partial_{zz}\Omega+\partial_{rr}\Omega- \frac{\partial_{r}\Omega}{r}\right)+4 r\partial_{r}\Omega. \label{EqForOmega}
\end{align}
In general, Eq. (\ref{EqForOmega}) for $\Omega$ is not needed if we can invert Eq. (\ref{SolV}). In fact, as discussed in Appendix (\ref{Solgenapp}) it possible to obtain a general solution for $\psi$, given by Eq. (\ref{eq:ax8}).
The function $\Phi$ and $\psi$ are completely integrated out, and the constant of motion $p_\phi$ is an arbitrary function of $\Omega$. 
{In addition, we note that the density approaches zero at large $r$ if 
\begin{align}
    \frac{ (\partial_r\Omega) ^2+(\partial_z\Omega) ^2}{8 \pi  G r^2} \rightarrow 0
\end{align}
sufficiently rapidly in the limit $r \rightarrow \infty$.}

\subsection{The effect of $p_\phi(\Omega)$}  \label{ssec:pphi}

We can build two tensors to characterize the first order geometric properties of the source fluid. As it is well known these are  the vorticity tensor $\mathbf{\Omega}$ and the deformations tensor $\mathbf{K}$ (Born's tensor) \cite{Ruggiero:2023ker}, defined as
\begin{align}
   \mathbf{\Omega}= du, \quad
   \mathbf{K}= \mathcal{L}_{u}(g).
\end{align}
We remember that $u$ is the four velocity of the fluid (\ref{eq:defv}) and $\mathcal{L}_{u}$ is the Lie derivative with respect to the vector field  $u$. In coordinates these two tensors read
\begin{align}
{\Omega}_{\mu\nu}&=\frac{\partial u_\mu}{\partial x^{\nu}}- \frac{\partial u_\nu}{\partial x^{\mu}},\\
{K}_{\mu\nu}&=\nabla_\mu u_{\nu}+\nabla_\nu u_{\mu}.
\end{align}
In our case, in the coordinates defined in Eq. (\ref{MFD}) and at the leading order we have
\begin{align}
\mathbf{\Omega}&=p_{\phi}'(\Omega) \left[d\Omega \wedge (d\phi-\Omega dt)\right]=\left[dp_{\phi}(\Omega) \wedge (d\phi-\Omega dt)\right], \\
\mathbf{K}&=r^2 \left[d\Omega \otimes (d\phi-\Omega dt)\right].
\end{align}
We note that the contraction of these tensors with $u$ gives zero, i.e. these tensors are purely spatial. The same result can be obtained performing the pull-back of these forms along the curve defined by the dust trajectory
\begin{align}
    \Omega= \frac{d\phi}{dt} \implies 
    d\phi=\Omega dt,
\end{align}
as in Eq. (\ref{eq:fluidvel}). We see that the information about the freedom of the function $p_\phi$ is in the antisymmetric part, which is the vorticity of the system.

\section{Mass and Angular momentum of the system}  \label{sec:komar}

To better investigate the physical properties of the system under consideration, we  study its  Komar integrals. To do this, we use spherical coordinates $(R,\theta)$ which are related to cylindrical ones $(r,z)$  in the usual way
\begin{align}
     r= R \sin{\theta}, \quad z= R \cos{\theta}.
\end{align}
The Komar mass is given by
\begin{align}
    M= \frac{c}{4\pi G} \int_{\partial \Sigma} d^2x \sqrt{\gamma^{(2)}} n_\mu \sigma_\nu \nabla^\mu \xi^\nu,
\end{align}
where $\partial \Sigma$ is a surface $(\theta,\phi)$ far away from the source.  The vector $\xi^\nu$ is the Killing vector $(\partial_t)^\nu$, while 
\begin{align}
    \sigma= dr, \quad
    n= c dt,
\end{align}
asymptotically for large $R$. Therefore, we find
\begin{align}
    n_\mu \sigma_\nu \nabla^\mu \xi^\nu= n_\mu \sigma_\nu g^{\mu\rho} \nabla_\rho \xi^\nu= n_0 \sigma_\nu g^{00}\nabla_0 \xi^\nu=-c\, \sigma_\nu \Gamma_{0\sigma}^{\,\,\,\nu}=c\,\Gamma_{00}^{\,\,\,R},
\end{align}
where we used that asymptotically $g^{00} \rightarrow -1,
    g_{RR} \rightarrow 1$.
In our case $\Gamma_{00}^{\,\,\,R}=-\frac{1}{2} g_{00,R}$, then
\begin{align}
    M=- \lim_{R\rightarrow \infty} \frac{R^2}{4\pi G} \int^{2\pi}_0 d\phi \int_0^\pi d\theta \sin\theta \left[\partial_R \Phi +\frac{\psi}{R \sin^2 \theta} \partial_R\frac{\psi}{R} \right].\label{MK}
\end{align}
For what concerns the angular momentum we find
\begin{align}
    J=-\frac{c^2}{8\pi G} \int_{\partial \Sigma} d^2x \sqrt{\gamma^{(2)}} n_\mu \sigma_\nu \nabla^\mu \xi^\nu, 
\end{align}
where this time $\xi^\nu=(\partial_\phi)^\nu$. In this case we have
\begin{align}
    n_\mu \sigma_\nu \nabla^\mu \xi^\nu= c\, \left(g^{00}\Gamma_{0\phi}^{\,\,\,R}+g^{0\phi}\Gamma_{\phi\phi}^{\,\,\,R}\right),  
\end{align}
where for at the leading order
\begin{align}
&\Gamma_{0\phi}^{\,\,\,R}=\frac{1}{2c} \partial_R \psi,\\
&\Gamma_{\phi\phi}^{\,\,\,R}=-R \sin^2\theta, \\
&g^{0\phi}= - \frac{\psi}{cR^2 \sin^2\theta}.
\end{align}
Putting these results together we end up with 
\begin{align}
    J=-\lim_{R\rightarrow \infty}\frac{c^2 R^2}{8\pi G} \int^{2\pi}_0 d\phi \int_0^\pi d\theta \sin\theta \left[-\frac{1}{2}\partial_R \psi +\frac{\psi}{R} \right]. \label{JK}
\end{align}
An interesting possibility to have a finite value for $J$ and a negligible contribution of these $\psi$ in the Komar mass $(\ref{MK})$ is given by 
\begin{align}
    \psi= 2\sum_{i} a_i G M \frac{r^2}{(r^2+(z-z_i)^2)^{3/2}}=2 \sum_{i} a_i G M \frac{\sin^2{\theta}}{R(1-\frac{2\cos\theta z_i}{R}+\frac{z_i^2}{R^2})^{3/2}},
\end{align}
where $a_i$ and $z_i$ are finite constants. This sum for $\psi$ is also a solutions of the equation of motion (\ref{Sgmsource1}) as showed in appendix \ref{Solgenapp}, moreover this is the Kerr angular profile at low energy. In fact, {using} this inside the angular momentum (\ref{JK}) we get
\begin{align}
    J= c^2 M \sum_i a_i,
\end{align}
which is proportional to the sum of different Kerr black holes angular momentum.
According to what we have seen in Section (\ref{sec:sol}), we know  how to write down the relation between $\psi$ and $V$. 
To keep the discussion simple, let us take the solution to be in the form
\begin{align}
     \psi= r V+ \alpha \frac{V^d}{r^d},
     \label{GenericTerm}
\end{align}
where $\alpha$ and $d$ are generic constants. In the following we exclude the case $d=-1$ that will be studied later.
Let us use this expression in Eqs. (\ref{SGMrr}) and (\ref{SGMzz}) (or in Eq. (\ref{Newtpot})) to obtain the potential
\begin{align}
   \Phi_d= -\frac{a r^{-2 (d+1)} V^d \left[a (d+1) V^d+2 V r^{d+1}\right]}{2 (d+1)}+ \text{constants}. \label{SolutionP}
\end{align}
The $r-$derivative is
\begin{align}
   \partial_r\Phi_d= \alpha  \,r^{-2 d-3} V^{d-1} \left[\alpha  V^d \left((d+1) V-d\, r\,
   \partial_r V\right)-V r^{d+1} \left(r \partial_r V-V\right)\right] \label{Derivativephi}
\end{align}
The density can  be found using Eq. (\ref{densityred}). For only one element of the series Eq. (\ref{GenericTerm}) we have for the density
\begin{align}
    \rho=\frac{ (\partial_r\Omega) ^2+(\partial_z\Omega) ^2}{8 \pi  G r^2} \left[ \alpha^2 d^2 \frac{V^{2 d-2}}{r^{2 d-2}}-r^4 \right],
\end{align}
we notice the positivity condition
\begin{align}
     \alpha d\,  V^{ d-1}-r^{d+1} >0, \label{Positivitycond}
\end{align}
{The above condition, for $d\neq1$ and $d\neq-1$, is}  an interesting relation between the size of the source and the asymptotic velocity. This condition of positivity tells us that if we vary the source size $R_S$ we expect the asymptotic velocity $V_\infty$ to vary as
\begin{align}
    V_{\infty}\propto R_S^{\frac{d+1}{d-1}}.
\end{align}
If we plug this into the above equations (\ref{Derivativephi}) and (\ref{MK}) we get
\begin{align}
    M&\sim R^2 \partial_R \Phi \propto R_S^{\frac{4}{d-1}+3} \propto V_{\infty}^{3-\frac{2}{d+1}},
\end{align}
{where the last part of this relation can be used only when $d\neq1$ and $d\neq-1$, according to the above remark.}
We see that for $d\gg0$ this asymptotically approaches to $M \propto V^3$, while for $-d\gg0$ it rapidly converges to a value between 3 and 4. The case $M\propto V^4$ is obtained for $d=-3$. 
This result is reminiscent of a Tully-Fisher-like relation, directly derived from the Komar mass of the system and the requirement of positive energy density. It is striking that such natural ingredients led to a direct relation between the asymptotic velocity and the total mass.

\section{Examples of solutions} \label{sec:exsolutions}

The dynamics of the system, together with its gravitational field, is self-consistently described in the low-energy limit by Eqs. (\ref{SGMzz})-(\ref{Sgmsource2}): the latter are a system in the variables $(r,z)$ and with four unknowns functions: $\Phi,\psi, \Omega,\rho$.  The Newtonian potential $\Phi$ fulfills Eq. (\ref{Newtpot});
on the other hand, as discussed in Appendix (\ref{Solgenapp}), the solution for $\psi$ can be written in equivalent forms
\begin{align}
 \psi=\int\limits_0^\infty d\lambda \cos{(\lambda z)}(r\lambda) A(\lambda)K_1(\lambda r)\, 
 \end{align}
 or,
 \begin{align}
 \psi=\int\limits_{-\infty}^\infty d\zeta \left[C(\zeta)\frac{r^2}{((z+\zeta)^2+r^2)^{3/2}} \right],
\end{align}
where $A(\lambda)$ and $C(\lambda)$ are generic spectra. After selecting the spectrum, the angular velocity $\Omega= r V$ in principle can be calculated using Eq. (\ref{SolV}). Finally, the density $\rho$ is obtained simply from the derivatives of the above functions, as shown in Eq. (\ref{Sgmsource2}). If we want to start from a given density, then the spectrums $A(\lambda)$ and $C(\lambda)$ must be taken accordingly.

To provide some explicit analytical examples, let us consider
\begin{align}
    \psi= r^2 \Omega+ \alpha \Omega^d,
\end{align}
where $\alpha$ and $d$ are generic constants. Of course we note that we can write an infinite sum of such solutions
\begin{align}
    \psi= r^2 \Omega+ \int_0^\infty K(\lambda)  \, \Omega^\lambda d\lambda,
\end{align}
where $K(\lambda)$ is a generic spectrum. 

We present explicit solutions by taking specific values of $d$ which amounts to setting $K(\lambda)= \alpha\, \delta(d-\lambda)$.

\subsection{Rigid rotation}
If $\Omega$ is constant, we cannot write Eq. (\ref{SolV}). In this case we have
\begin{align}
    V= \Omega_0 r,
\end{align}
where $\Omega_0$ is a constant. If we proceed to integrate the equations we get the same solutions discussed by \citet{Balasin:2006cg,Cooperstock:1993en}.

\subsection{Case with $d=-1$ } \label{AsimpFlat}
When $d=-1$, we then have the following solution 
\begin{align}
    \Omega(r,z)= r V(r,z)&= \frac{\psi (r,z)\mp \sqrt{ \psi (r,z)^2+ 4 V^2_\infty r^2} }{2  }, \\
    \Phi(r,z)&=\frac{V_\infty^2}{2 V(r,z)^2} \left(V(r,z)^2-V_\infty^2\right)-V_\infty^2 \log \left(\frac{V_\infty
   }{V(r,z)}\right)-V_\infty^2 \log \left(\frac{r}{r_0}\right),\\
    \psi(r,z)&=r \frac{ V^2-V_\infty^2}{V}, \\
   \rho(r,z)&= \frac{\left[r^2 (\partial_z V)^2+\left(V-r
   \partial_r V\right)^2\right] }{8 \pi  G
   r^2 } \frac{\left(V^4_\infty-V^4\right)}{V^4}.
\end{align}
where the constant $V_{\infty}$ is the maximum dust velocity and $r_0$ is a constant. We see that if at large distances the function $\psi$ goes to zero faster than $V$, we get an asymptotic flat velocity profile. There are also other possibilities to obtain a flat velocity profile. 
We see that the  potential $\Phi$ exhibits  a $\log(r)$ divergence; this divergence can be resolved by carefully choosing the function $\psi$. However, since the system we are dealing with may have a large but finite extension, at some distance the solution must be truncated and joined to a vacuum solution. Then, we got an infinite class of solutions with an asymptotic flat velocity profile. The peculiar feature is that the maximum velocity is reached asymptotically.
To give an example let us take the seed 
\begin{align}
    \psi=\pm \frac{ r^2}{(z^2+r^2)^{3/2}}. \label{KerrAng}
\end{align}
The density on the equatorial plane becomes
\begin{align}
   \rho(r,0)= \frac{\left(\sqrt{4 r^4 V_{\infty}^2+1}-2\right)^2}{8 \pi  G r^6
   \sqrt{4 r^4 V_{\infty}^2+1}}, 
\end{align}
which asymptotically goes as $r^{-4}$.
\begin{figure}[h!]
    \centering
\includegraphics[width=0.75\textwidth]{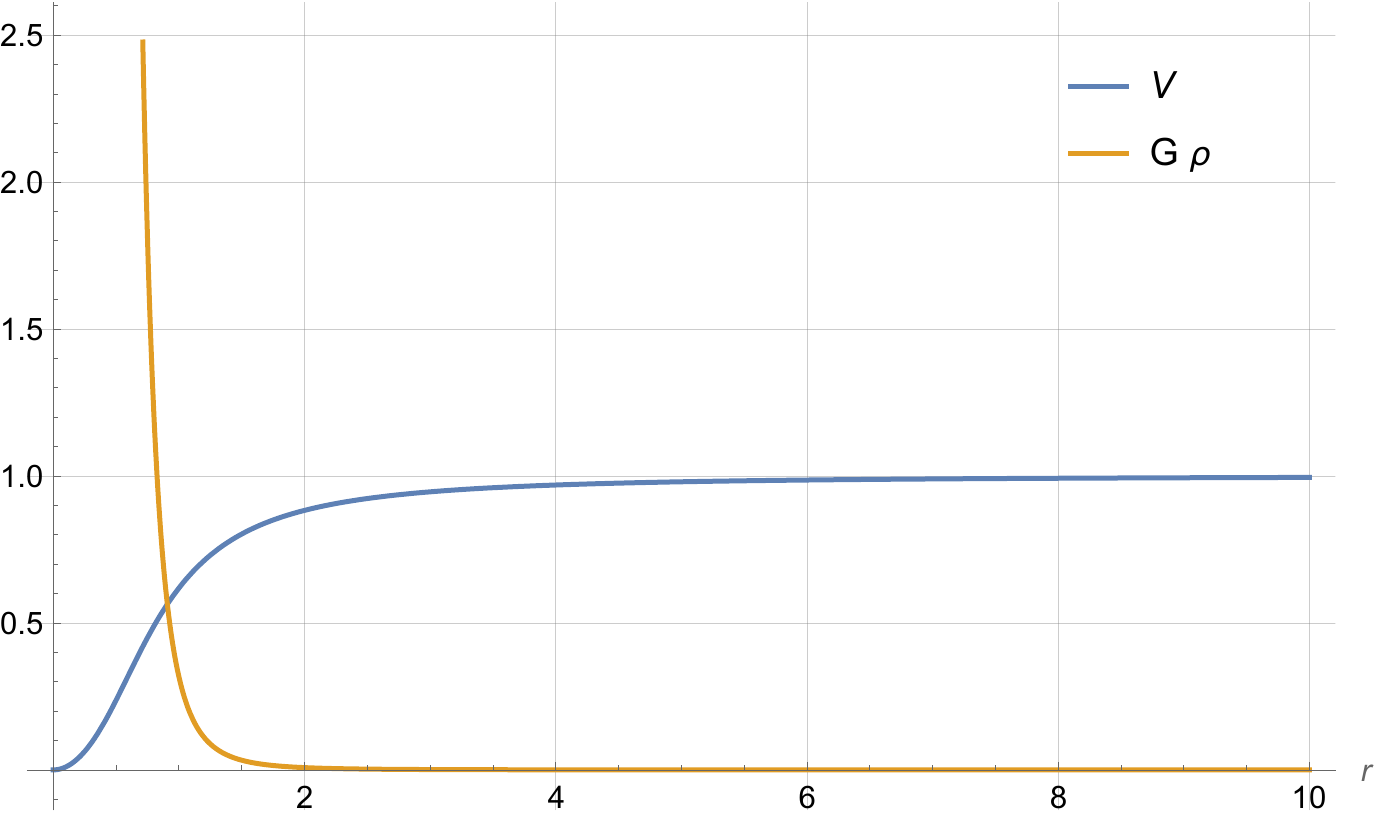}  
    \caption{Density for $V_{\infty}=1$.}
    \label{fig:sample-image}
\end{figure}

\subsection{Case with $d=1$ }
When $d=1$ we have
\begin{align}
    \psi= r^2 \Omega+ k\, r_0^2\, \Omega, \qquad \Phi= \frac{ k\, r_0^2 \left(r^2+ k\, r_0^2\right) \Omega (r,z)^2}{2 r^2},
\end{align}
where $k$ is a sign and $r_0^2$ is a constant. In this case it is easy to invert this equation and we find that the general solution for $\Omega$ is given by
\begin{align}
    \Omega(r,z)=\frac{1}{r^2+ k r_0^2}\int\limits_0^\infty d\lambda \cos{(\lambda z)}(r\lambda)\left[A(\lambda)K_1(\lambda r)\right],
\end{align}
after we threw away the term proportional to $I_1(\lambda r)${(which is the diverging part of the solution expressed in terms of Bessel functions, as we discuss in Appendix \ref{Solgenapp}}), to avoid singularities.
Now the density is completely given after we fix the spectrum $A(\lambda)$
\begin{align}
 \rho=\frac{\left(\partial_r\Omega\right)^2+\left(\partial_z\Omega\right)^2}{8 \pi  G r^2} \left(r_0^4-r^4\right).
\end{align}
Also in this case it is possible to have a flat profile. We want to elucidate in this case how to fix $A(\lambda)$ based on the density profile. 
Plugging the solution for $\Omega$ inside the density we get in the equatorial plane $(z=0)$
\begin{align}
 2 \sqrt{2 \pi G \rho(r,0)} \frac{\left(k r_0^2+r^2\right)^{3/2}}{\sqrt{ k r_0^2-r^2}}=  \int_0^{\infty }\lambda  A(\lambda ) \left[\lambda \left(k r_0^2+r^2\right) 
    K_0(r \lambda )+2 r K_1(r \lambda )\right] d\lambda.
\end{align}
In general this can be inverted but it is a complicate task. For the purpose of this paper let's just consider a region where $r_0 \gg r $, where also the first term is dominant respect to the second one in the r.h.s. If we multiply both sides by $r K_0(r \lambda' )$ and integrate over $r$ we get
\begin{align}
 A(\lambda)= \frac{2}{\lambda} \int_0^{\infty} \sqrt{2 \pi G \rho(r,0)} r K_0(r \lambda ) dr,
\end{align}
where we made used of the orthogonality properties between Bessels' functions. Now we can insert the observed energy density and obtain the respective spectral density $A(\lambda)$.

\section{General solution for zero energy density } \label{vacuum}
\begin{figure}[h!]
    \centering   \includegraphics[width=0.85\textwidth]{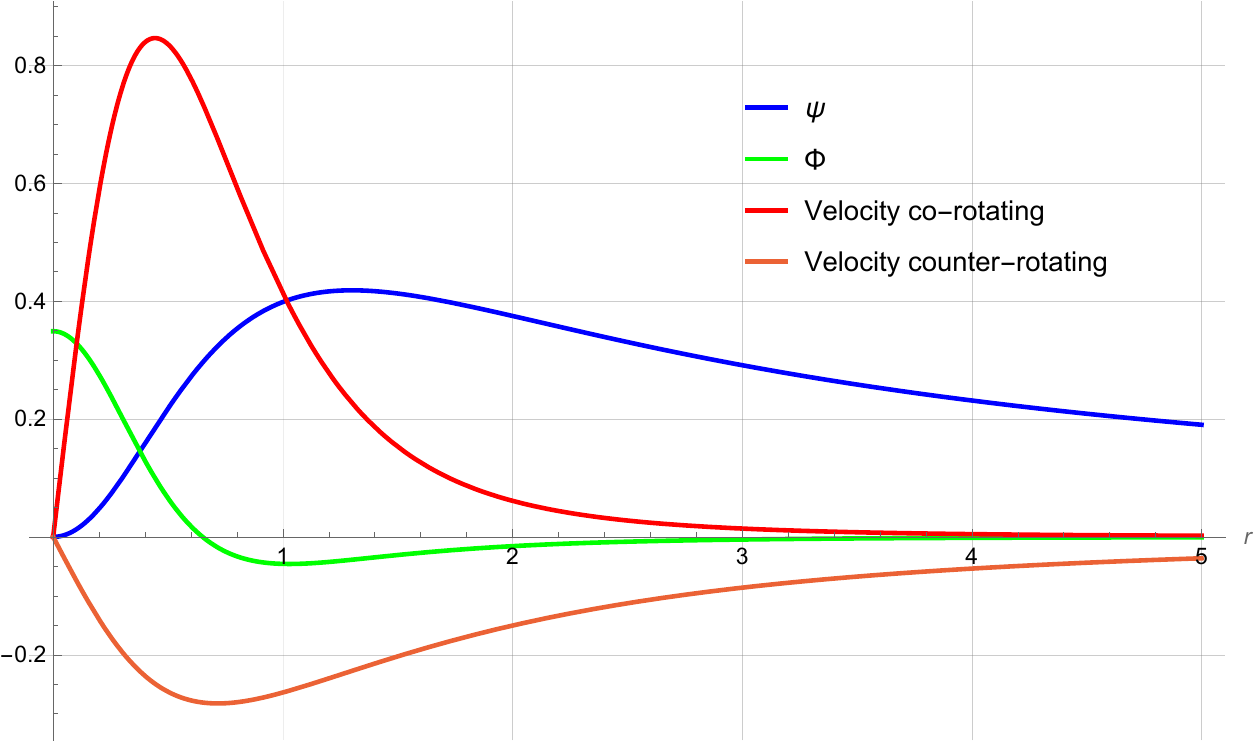} 
    \caption{A vacuum solution, in the equatorial plane $z=0$.  }
    \label{fig:vacuum}
\end{figure}

It is interesting to discuss the case of zero energy density. These solutions are relevant when we investigate the behaviour far away from the source, where the energy density is negligible. Another domain of applicability of the solution is in a region where there is no source at all. Therefore, we are going to solve the following equations
\begin{align}
\partial_{rr} \psi+ \partial_{zz} \psi- \frac{\partial_r \psi}{r}&=0, \\
    \nabla^2 \Phi+ \frac{(\partial_{z}\psi)^2+(\partial_r\psi-2 \frac{\psi}{r})^2}{2 r^2}&=0.
\end{align}
To start with a simple example, let us assume again that the angular function $\psi$ is given by the low-energy profile
\begin{align}
   \psi=a_1 \frac{ r^2}{((z-z_1)^2+r^2)^{3/2}},
\end{align}
then a zero energy density $\rho$ is obtained by direct integration of $\Phi$. This gives
\begin{align}
   \Phi=\frac{a_1^2 \left((z-z_1)^2-2 r^2\right)}{4 \left(r^2+(z-z_1)^2\right)^3}+\frac{a_0}{
   \sqrt{r^2+z^2}}.
\end{align}
We note that the first term is like a quadrupole moment, while the second exhibits the expected Newtonian behavior. 
The radial geodesic equation gives two solutions, which exhibit the profiles showed in Figure \ref{fig:vacuum}, with $a_1=1$, $a_0=0$ and $z_1=0.92$ in the equatorial plane.\\

In general we have an infinite sum of such terms in the angular potential 
\begin{align}
   \psi= \sum^n_i a_i \frac{ r^2}{((z-z_i)^2+r^2)^{3/2}}.
\end{align}
The solution for the potential can be written as well as a series and reads
\begin{align}
   \Phi=\frac{a_0}{
   \sqrt{r^2+z^2}}+ \sum^n_{i=1} \sum^n_{j=1 }\frac{a_i a_j \left((z-z_i) (z-z_j)-2 r^2\right)}{4 \left(r^2+(z-z_i)^2\right)^{3/2}\left(r^2+(z-z_j)^2\right)^{3/2}}.
\end{align}
In the continuous limit $n \rightarrow \infty$ they become
\begin{align}
    \psi&= \int_{-\infty}^{\infty}  \frac{ C(\zeta) \, r^2}{((z-\zeta)^2+r^2)^{3/2}} d\zeta, \\
    \Phi&=\frac{a_0}{
   \sqrt{r^2+z^2}}+ \int_{-\infty}^{\infty} \int_{-\infty}^{\infty} \frac{C(\zeta) C(\zeta')\left((z-\zeta) (z-\zeta')-2 r^2\right)}{4 \left(r^2+(z-\zeta)^2\right)^{3/2}\left(r^2+(z-\zeta')^2\right)^{3/2}} d\zeta d\zeta',
\end{align}
which is the general solution and depends on the choice of a generic spectrum $C(\zeta)$. The spectrum can be fixed to obtain particular orbits for a particle moving in this background.
In this solution we note that the ``source'' of the Newtonian potential is the energy-momentum of the rotation of spacetime determined by $\psi$.

{The difference between the rotating and co-rotating velocity fields can be a signature of the presence of strong angular terms in such systems. In our approximation, it reads
\begin{align}
   V_{+}-V_{-}= \sqrt{-4 r \phi ^{(1,0)}(r,z)+\left(\partial_r \psi(r,z)-\frac{2 \psi (r,z)}{r}\right)^2},
\end{align}
the first is a Newtonian term including $\phi$. More interesting is to have a look at the second term, which is proportional to the energy density "generated" by the presence of the dragging term. In the equatorial plane, due to the symmetry, we can assume $\partial_z \psi=0$, therefore the density in eq. ({\ref{eq:rhopsi}}) becomes exactly the second term of the difference between the two velocities.}

\section{Conclusions} \label{sec:conc} 

In this work, we focused on axisymmetric, stationary systems in General Relativity. Specifically, we considered rotating dust as the source of the gravitational field and investigated the low-energy limits of the solutions to Einstein's equations. The approach we adopted is  new, since we started from the low-energy limit of the metric and then consistently solved Einstein's equations; in a previous approach the low-energy limit was obtained by performing an expansion of the exact solutions of Einstein's equations. We proved that in both cases the results are the same. 

We emphasized that there is a low-energy limit, which we called strong gravitomagnetism, where the dragging terms  are determined by the solution of a homogeneous Grad-Shafranov equation which does not depend on the sources of the gravitational field. 
Consequently, these terms are unaffected by the expansion of the source in powers of $V/c$ and are therefore not negligible compared to the Newtonian terms. In other words, we have demonstrated that, in the low-energy limit, the differences between the dynamics of General Relativity and the Newtonian one persist. The existence of such systems fundamentally relies on the presence of these terms. Without them, the system would exhibit cylindrical symmetry, meaning it would extend infinitely along the axis of symmetry. This is precisely the case in Newtonian gravity, where it is impossible to construct a finite, stationary, rotating system made of dust with axial symmetry. In essence, the system under consideration is unique because it lacks a Newtonian counterpart. 

After pointing out the peculiarities of such a system and analyzing its physical properties thanks to the use of the Komar integrals for its mass and angular momentum, we discussed an approach to a self-consistent solution of its equation of motions in the strong gravitomagnetic case. In particular, we outlined a procedure to solve the equations that self-consistently describe both the dynamics of the system and its gravitational field. In addition, we gave the more general solution and an example of the solution of the equations for the  vacuum case. To emphasize the key characteristics of these solutions, we considered some particular cases and pointed that, due to its deeply non-Newtonian nature, the dynamics of the system is completely different with respect to what is expected on the basis of a classical (i.e. non-relativistic) analysis.  For instance, we showed that the velocity profile describing the motion of the sources can be asymptotically flat.  Regarding this, we know that the flatness of rotation curves is a very important issue in the study of galactic dynamics and one of the evidences for the presence of dark matter:{in particular, recent results \cite{Mistele:2024hfh} suggest that these curves remain flat for hunderds of kpc, thus extending previous observations.} So, it is not a surprise that  models of self-gravitating systems constituted by dust, in stationary rotation with axial symmetry were considered as relativistic models for galaxies  \cite{Balasin:2006cg,Cooperstock:1993en,crosta2020testing,Astesiano:2021ren,Astesiano:2022gph,Astesiano:2022ozl,Astesiano:2022ghr,Ruggiero:2023geh,10.1093/mnras/stae855,Re:2024qco,Galoppo:2024ttc,Galoppo:2024mfw}. So, apart from the mathematical interest in these systems, the open question remains whether they can be considered models for real astrophysical objects. We believe this question holds potential relevance for research in relativistic astrophysics.

\section*{Acknowledgments}
The work of D.A. is supported in part by Icelandic Research Fund grant 228952-052. MLR thanks the support of the Gruppo Nazionale per la Fisica Matematica (GNFM).
The authors thank Fri{\dh}rik Freyr Gautason for important suggestions on a preliminary version of this work; in addition,  the authors thank Sergio Cacciatori, Federico Re, Marco Galoppo, David Wiltshire for stimulating discussions.

\newpage

\appendix
\section{Solutions of the homogenous equation } \label{Solgenapp}

Here we explicitly solve the homogeneous equation (\ref{Sgmsource1}) which we rewrite  for convenience
\begin{align}
    \partial_{zz}\psi+\partial_{rr}\psi-\frac{1}{r }\partial_r \psi=0.
\end{align}
We remember that $\psi$ determines the dragging effects that are relevant in the low-energy limit and are greater than the usual ones considered in the standard gravitomagnetic approach. The $r-$ dependence of the $\psi$ function can be fixed once we choose the $z-$ behaviour of the same function. In fact, if we employ the Ansatz
\begin{align}
  \psi={\mathcal R}(r){\mathcal Z}(z),
\end{align}
we arrive at 
\begin{align}
    {\mathcal Z}_{zz}=k{\mathcal Z},
\end{align}
with the separation constant $k\in\mathbb{R}$. At this point we want to impose reflection symmetry $\psi(r,z)=\psi(r,-z)$.
We now have two possibilities. The first one is to assume a positive value for $k:=u^2$ (with $u\in\mathbb{R}_0^+$) and the fall off behavior at infinity can be achieved by employing the non-smooth modes ${\mathcal Z}(k,z)=e^{-\sqrt{k}|z|}$, which satisfy \eqref{Sgmsource1} for $|z|>0$ only. This was done in \cite{Cooperstock:2005qw}. Consequently, there are sources localized at $z=0$. Let us find the solution in this case. 
We start from 
\begin{align}
    \tilde{U}(u,z):= \int^\infty_0  \psi(r,z) J_1(ru) dr,
\end{align}
where $J_m(ru)$ is the Bessel function of the order $m$. Then, the second derivative with respect to $z$ is
\begin{align}
    \partial_{zz} \tilde{U}=\int^\infty_0 \partial_{zz} \psi(r,z) \, J_2(ru) dr=- \int^\infty_0 \left( \partial_{rr}\psi-\frac{1}{r}\partial_r \psi \right) J_1(ru) dr,
\end{align}
after a partial integration we have
\begin{align}
    \partial_{zz} \tilde{U}= u \int^\infty_0   \partial_r \psi\, J_0(r u) dr,
\end{align}
after another partial integration we get
\begin{align}
    \partial_{zz} \tilde{U}=u^2 \int^\infty_0 \psi \,  J_1(ru) dr = u^2 \Tilde{U}.
\end{align}
This implies that the required form is $\tilde{U}(u,z)= A(u) e^{-u|z|}+B(u)e^{u |z|}$. The asymptotic boundary condition 
\begin{align}
    \lim_{z\rightarrow\infty} \psi=0,
\end{align}
implies $B(u)=0$. We have obtained
\begin{align}
     \tilde{U}(u,z)=A(u)e^{-u|z|}= \int^\infty_0  \psi(r,z) J_1(ru) dr. 
\end{align}
To obtain the final expression for  $\psi$ now it suffices to invert the relation using the Hankel transform. The result is
\begin{align}
    \psi(r,z)= \int^\infty_0 r u A(u) e^{-u|z|} J_1(ru) du,
\end{align}
due to the arbitrariety of $A(u)$ we can redefine a new function using $\tilde{\psi}(u):=u A(u)$:
\begin{align}
    \psi(r,z)=\int^\infty_0 r e^{-\lambda |z|}\tilde{\psi}(\lambda) J_1(\lambda r) d\lambda, \label{psi}
\end{align}
\medskip

Another possibility is  ${\mathcal Z}(k,z)=\cosh{(\sqrt{k}z)}$ as in \citet{Balasin:2006cg}. The parameter $k$ cannot be positive then, otherwise these modes diverge exponentially for $|z|\to\infty$, which is an unphysical behavior. In this case we assume $k=-\lambda^2$ with $\lambda\in\mathbb{R}_0^+$ yielding the modes ${\cal Z}(\lambda,z)=\cos{(\lambda z)}$.
The solution can be found with the same procedure as in the first case. The result is given by 
\begin{equation}
  \label{eq:ax8}
  \psi(r,z)=\int\limits_0^\infty d\lambda \cos{(\lambda z)}(r\lambda)\left[A(\lambda)K_1(\lambda r)+B(\lambda)I_1(\lambda r)\right]\,.
\end{equation}
The functions $I_1$ and $K_1$ are modified Bessel functions of the first and second kind, respectively. We note that $I_1$ blows up exponentially for large values of $r$ which is unphysical. Therefore, we set $B(\lambda)=0$. The function $K_1$ falls off exponentially for large values of $r$ and diverges like $1/r$ near $r=0$. However, this divergence is compensated by a linear prefactor, so the integrand is well defined for any sufficiently regular $A(\lambda)$.
    We can write down the result in a more interesting fashion. We perform first a Fourier transformation,\begin{equation}
  \label{eq:ax25}
  A(\lambda)=\frac{2}{\pi}\int\limits_0^\infty d x \,C(x)\cos{(\lambda x)}\,,
\end{equation}
where $A(\lambda)$ is determined in terms of a (Fourier) transformed spectral density $C(x)$.
After using the following property of Bessel functions
\begin{equation}
  \label{eq:ax23}
  \int\limits_0^\infty d x x K_1(x) \cos{(\frac{a x}{r})} = \frac{\pi}{2}\frac{r^3}{(a^2+r^2)^{3/2}}\,,
\end{equation}
the form of $\psi$ becomes 
\begin{align}
  \psi(r,z) 
= \frac{1}{2}\int\limits_0^\infty d\zeta \left[C(\zeta)\frac{r^2}{((z+\zeta)^2+r^2)^{3/2}} 
+C(\zeta)\frac{r^2}{((z-\zeta)^2+r^2)^{3/2}}\right].
\label{Gensol}
\end{align}
We recognize this to be in  form of Eq. (\ref{KerrAng}): we are basically performing an integral over all the possible forms of this kind with different frequencies. In fact, in particular we have the following sum
\begin{align}
    \psi(r,z)=r^2 \left( \frac{c_1}{((z+d_1)^2+r^2)^{3/2}}+\frac{c_2}{((z+d_2)^2+r^2)^{3/2}}+... \right), \label{SumKerr}
\end{align}
which is a solution.

\section{Strong gravitomagnetism with pressure terms}  \label{Addition of pressure}

Here we are concerned with an axially symmetric and stationary rotating system, where the sources are determined by the energy momentum tensor for a perfect fluid 
\begin{align}
    T_{\mu\nu}=(\rho+\frac{p}{c^2})  u_\mu u_\nu+p g_{\mu\nu},
\end{align}

with $\mu\nu=0,1,2,3$. The trace reversed Einstein equations and the Ricci tensor are
\begin{align}
    R_{\mu\nu}&= \frac{8 \pi G}{ c^4} \left[\frac{(\rho-\frac{p}{c^2})g_{\mu\nu}}{2}+\left(\rho+\frac{p}{c^2}\right)u_\mu u_\nu\right], \\
    R&=\frac{8 \pi G}{ c^4} (\rho-3 \frac{p^2}{c^2}).
\end{align}
Since $p/c^2$ is $O(c^{-2})$ at our order of approximation it does not back react on the geometry and the equations for the sources are the same as the ones that we have discussed above. For example in the strong gravitomagnetic plus pressure case we still have Eqs. (\ref{Sgmsource1}) and (\ref{Sgmsource2}). These pressure terms give a non trivial contribution when we impose
\begin{align}
    \nabla_\nu T^{\mu\nu}=0.
\end{align}
Explicitly it gives
\begin{align}
    (\rho+\frac{p}{c^2})_{;\nu} u^\mu u^\nu+(\rho+\frac{p}{c^2}) u^\mu u^\nu_{;\nu}+(\rho+\frac{p}{c^2}) u^\mu_{;\nu} u^\nu+p^{,\mu}=0.
\end{align}
We are using the usual conventions $\nabla_\mu A^\nu:=A^\nu_{;\mu}$ and $\partial_\mu A^\nu:=A^\nu_{,\mu}$. The projection of this equation along the four velocity flow is
\begin{align}
    u_\mu\nabla_\nu T^{\mu\nu}=0 \implies c^2 (\rho+\frac{p}{c^2}) u^\nu_{;\nu}=- u^\nu c^2 \rho_{,\nu},
\end{align}
where we used $g_{\mu\nu}u^\mu u^\nu=-c^2$. The space projector is defined by $P_{\mu\nu}=g_{\mu\nu}+u_\mu u_\nu$ and gives
\begin{align}
  P^\alpha_{\mu}\nabla_\nu T^{\mu\nu} =0 \implies  (\rho+\frac{p}{c^2}) u^\nu u_{\mu;\nu}=- \left(p_{,\mu}+u_\mu u^\nu p_{,\nu}\right).
\end{align}
We can also rewrite it in the following way
\begin{align}
    (\rho+\frac{p}{c^2}) u^\nu u^\mu_{;\nu}=- \left(g^{\mu\nu}p_{,\nu}+u^\mu u^\nu p_{,\nu}\right). \label{CovT}
\end{align}
In the strong gravitomagnetic limit the metric  and the four velocity of matter are
\begin{align}
     ds^2&= -c^2\left(1-\frac{2\Phi}{c^2}-\frac{\psi^2}{r^2c^2}\right)dt^2-2 \frac{\psi}{c} c dt d\phi +r^2\left(1+ \frac{2\Phi}{c^2}\right)d\phi^2+ e^{\Psi} \left(dr^2+dz^2\right),  \\
     u^t&=1+ \frac{\Phi}{c^2}+ \frac{v^2}{2c^2}, \quad u^\phi=\Omega=\frac{\psi}{r^2}+ \frac{v}{r},
\end{align}
such that $V= r \Omega$ is the velocity measured by asymptotic inertial observers. Then, in these coordinates Eq. (\ref{CovT}) gives non-trivial results on the $r$ and $z$ directions, which are respectively
\begin{align}
    (\rho+\frac{p}{c^2}) \left[(u^t)^2 \Gamma^r_{tt}+2 u^t u^\phi \Gamma^r_{t\phi}+(u^\phi)^2 \Gamma^r_{\phi\phi}\right]=-g^{rr} p_{,r},\\
    (\rho+\frac{p}{c^2}) \left[(u^t)^2 \Gamma^z_{tt}+2 u^t u^\phi \Gamma^z_{t\phi}+(u^\phi)^2 \Gamma^z_{\phi\phi}\right]=-g^{zz} p_{,z}.
\end{align}
At the leading order in $c$ we get 
\begin{align}
     \rho \left[-\frac{V}{r} \partial_z \psi+\partial_z \Phi+ \frac{\psi}{r^2} \partial_z \psi\right]&=p_r, \label{SGMzzp}\\
     \rho \left[-\frac{V}{r} \partial_r \psi+\partial_r \Phi+  \frac{V^2}{r}+ \frac{\psi}{r} \partial_r \frac{\psi}{r}\right]&=p_z\label{SGMrrp}\\
   \partial_{rr} \psi+ \partial_{zz} \psi- \frac{\partial_r \psi}{r}&=0, \label{Sgmsource1p}\\
    -\frac{1}{4\pi G} \left[\nabla^2 \Phi+ \frac{(\partial_{z}\psi)^2+(\partial_r\psi-2 \frac{\psi}{r})^2}{2 r^2}\right]&=\rho\label{Sgmsource2p}.
\end{align}
The gravitomagnetism with pressure is obtained when we neglect the $\psi^2$ terms. The Newtonian limit with pressure is recovered with $\psi \rightarrow 0$. We see that the source equations remain the same at the leading order when we add the pressure terms. Eq.(\ref{SGMrrp}) and (\ref{SGMzzp}) can be rewritten as
\begin{align}
    \rho \left(d\Phi+\frac{V^2}{r} dr\right)= dp, \\
    \rho \left(d\Phi+\frac{V^2}{r} dr- \frac{V}{r} d\psi\right)= dp,\\
    \rho \left(d\Phi+\frac{V^2}{r} dr- V r d(\frac{\psi}{r^2})\right)= dp, 
\end{align}
for the Newtonian, standard gravitomagnetic and strong gravitomagnetic cases, respectively.

\section{{Scalar-Vector-Tensor Analysis of the strong gravitomagnetic solution}}  \label{sec:scalar}

Here we analyze the solution (\ref{eq:defv}) to point out its physical degrees of freedom. To this end, we perform a Scalar-Vector-Tensor decomposition, which is a standard approach to study perturbations of Minksowski spacetime \cite{carroll2019spacetime}.

The metric (\ref{eq:defv}) can be written in Cartesian coordinates $(x^0=ct,x,y,z)$ by using the coordinate transformation $\phi=\arctan(y/x)$, $r=\sqrt{x^2+y^2}$. In general, the explicit expression of this metric depends on the function $\Psi$; the latter is defined by the line integrals (\ref{eq:defPSIr}),(\ref{eq:defPSIz}) which can be solved once that a specific function $\psi$ is chosen. To fix the ideas, we consider $\psi$ in the form (\ref{KerrAng})
\begin{align}
    \psi= \frac{m^2 r^2 c}{(z^2+r^2)^{3/2}}. \label{KerrAng1}
\end{align}
where we introduced the constant $m$ with the dimension of a length. We then obtain

\begin{align}
     ds^2&= -c^2\left(1-\frac{2\Phi}{c^2}-\frac{\psi^2}{(x^2+y^2)c^2}\right)dt^2-2 \frac{\psi}{c(x^2+y^2)} c dt (xdy-ydx)+\\
     &+\left( 1+ \frac{2\Phi}{c^2}\right)(dx^2+dy^2+dz^2)+\zeta dz^2
     +\frac{\zeta}{x^2+y^2}\left(x^2 dx^2+y^2 dy^2 +2xydxdy\right)
      \label{eq:svtmetric}
\end{align}
where
\begin{equation}
\zeta=\frac 9 8 \frac{m^4(x^2+y^2)^2}{(x^2+y^2+z^2)^4}
\end{equation}
From the expression (\ref{eq:svtmetric}) of the metric tensor $g_{\mu\nu}$, we define the $h_{\mu\nu}$ as the perturbation of the Minkowski tensor $\eta_{\mu\nu}$, $g_{\mu\nu}=\eta_{\mu\nu}+h_{\mu\nu}$. Accordingly, we have
\begin{align}
&h_{00}=\frac{2\Phi}{c^2}+\frac{\psi^2}{(x^2+y^2)c^2},\ h_{xx}=\frac{2\Phi}{c^2}+\frac{\zeta x^2}{x^2+y^2}, \ h_{yy}=\frac{2\Phi}{c^2}+\frac{\zeta y^2}{x^2+y^2}, \ h_{zz}=\frac{2\Phi}{c^2}+\zeta, \\
& h_{xy}=\frac{\zeta x y}{x^2+y^2},\ h_{0x}=\frac{1}{c}\frac{\psi}{x^2+y^2}y, h_{0y}=-\frac{1}{c}\frac{\psi}{x^2+y^2}x
\end{align}

The metric (\ref{eq:svtmetric}) can be written in the form \cite{carroll2019spacetime}

\begin{equation}
ds^2=-(1+2\sigma)+2w_idt dx^i+\left[(\left(1-2\tau \right)\delta_{ij}+2s_{ij} \right]dx^idx^j \label{eq:metricasvt}
\end{equation}
where Latin index refer to spatial components. In the above equation (\ref{eq:metricasvt}) we set
\begin{align}
&h_{00}=-2\sigma, \ h_{0i}=w_i, \ h_{ij}=2s_{ij}-2\tau \delta_{ij}    
\end{align}
with
\begin{align}
\tau=-\frac 1 6 \delta^{ij}h_ij, \ s_{ij}=\frac 1 2 \left(h_{ij}-\frac 1 3 \delta^{km}h_{km} \delta_{ij} \right)    
\end{align}
The above form of the metric (\ref{eq:metricasvt}) emphasizes the transformation properties of the perturbation tensor $h_{\mu\nu}$ under spatial rotations: in fact, the $00$ component is a scalar, the $0i$ is a three-vector, and the $ij$ components form a (symmetric) spatial tensor. In addition, $\tau$ contains the trace of $h_{ij}$, while $s_{ij}$ is traceless. 

In particular, we obtain
\begin{align}
&\sigma=-\frac 1 2 \left(\frac{2\Phi}{c^2}+\frac{\psi^2}{(x^2+y^2)c^2}\right), \ s_{xx}=\frac{\zeta}{2}\left(\frac{x^2}{x^2+y^2}-\frac 2 3 \right), \ s_{yy}=\frac{\zeta}{2}\left(\frac{y^2}{x^2+y^2}-\frac 2 3 \right) \\
& s_{xx}=\frac 1 6 \zeta,\  s_{xy}=\frac 1 2 \frac{\zeta x y}{x^2+y^2}, \ \tau=-\left(\frac{\Phi}{c^2}+\frac 1 3 \zeta \right), \   w_{x}=\frac{1}{c}\frac{\psi}{x^2+y^2}y, \ w_{y}=-\frac{1}{c}\frac{\psi}{x^2+y^2}x
\end{align}

We note that, differently with respect to the standard gravitomagnetic limit (see e.g. \citet{Ruggiero:2023ker,Astesiano:2022ozl}) where the $h_{00}$ term contains only the Newtonian or gravitoelectric potential, in this case  the scalar perturbation $\sigma$ is not decoupled from the vector perturbation, which is determined by the $\psi$ function.  


\bibliography{refs}

\end{document}